\pdfoutput=1

\documentclass{llncs}

\usepackage{latexsym}
\usepackage{times}
\usepackage{mathptm}
\usepackage{listings}
\usepackage{multirow}
\usepackage{rotating}

\usepackage{subfigure}
\usepackage{boxedminipage} 
\usepackage{url} 
\usepackage{AMMALanguages}
\newcommand{\DEBUG}{ON}

\newcommand{\AP}[1]{\ifthenelse{\equal{\DEBUG}{ON}}{\marginnote{Alfonso}{#1}}{}}
\newcommand{\DDR}[1]{\ifthenelse{\equal{\DEBUG}{ON}}{\marginnote{Davide}{#1}}{}}
\newcommand{\RL}[1]{\ifthenelse{\equal{\DEBUG}{ON}}{\marginnote{Antonio}{#1}}{}}
\newcommand{\NOTE}[1]{\ifthenelse{\equal{\DEBUG}{ON}}{\marginnote{NOTE}{#1}}{}}

\newcommand{\tag}[1]{{\small \texttt{<\kern-2pt[#1]\kern-2pt>}}\xspace}

\newcommand{\marginnote}[2]{\marginpar{\parbox{2cm}{\flushleft
 \tiny\textbf{#1}: #2}}}

\newcommand{\sound}{$\bullet$}
\newcommand{\unsound}{$\circ$}
\newcommand{\broken}{$\times$}
\newcommand{\sw}[1]{\ \ \begin{sideways}#1\end{sideways}\ \ \ }

\begin{document}

\title{\vspace{-42\in}%
Automated co-evolution of GMF editor models
}

\author{%
\vspace{-42\in}%
Davide Di Ruscio$^1$,
Ralf L\"ammel$^2$,
Alfonso Pierantonio$^1$}

\institute{%
{\small $^1$ Software Languages Team, Universit\"at Koblenz-Landau, Germany}\\
{\small $^2$ Computer Science Department, University of L'Aquila, Italy}}

\maketitle
\thispagestyle{empty}

\begin{abstract}
  \vspace{-66\in} The Eclipse Graphical Modeling (GMF) Framework
  provides the major approach for implementing visual languages on top
  of the Eclipse platform. GMF relies on a family of modeling
  languages to describe different aspects of the visual language and
  its implementation in an editor. GMF uses a model-driven approach to
  map the different GMF models to Java code. The framework, as it
  stands, provides very little support for evolution. In particular,
  there is no support for propagating changes from say the domain
  model (i.e., the abstract syntax of the visual language) to other
  models. We analyze the resulting co-evolution challenge, and we
  provide a transformation-based solution, say GMF model adapters,
  that serve the propagation of abstract-syntax changes based on the
  interpretation of difference models.
\end{abstract}



\vspace{-77\in}

\section{Introduction}
\label{sec:introduction}

\vspace{-33\in}

``The Eclipse Graphical Modeling Project (GMP) provides a set of
generative components and runtime infrastructures for developing
graphical editors based on EMF and GEF.''~\cite{GMF}. Arguably, GMF
defines the mainstream approach to graphical editor development within
the Eclipse platform. The approach heavily relies on metamodeling
(based on EMF~\cite{EMF1,EMF2}), model-to-model and model-to-code
transformations, and even some forms of code customization, subject to
round-tripping considerations.


\begin{figure}[h!]
\vspace{-42\in}
	\centering
		\includegraphics[width=.95\linewidth]{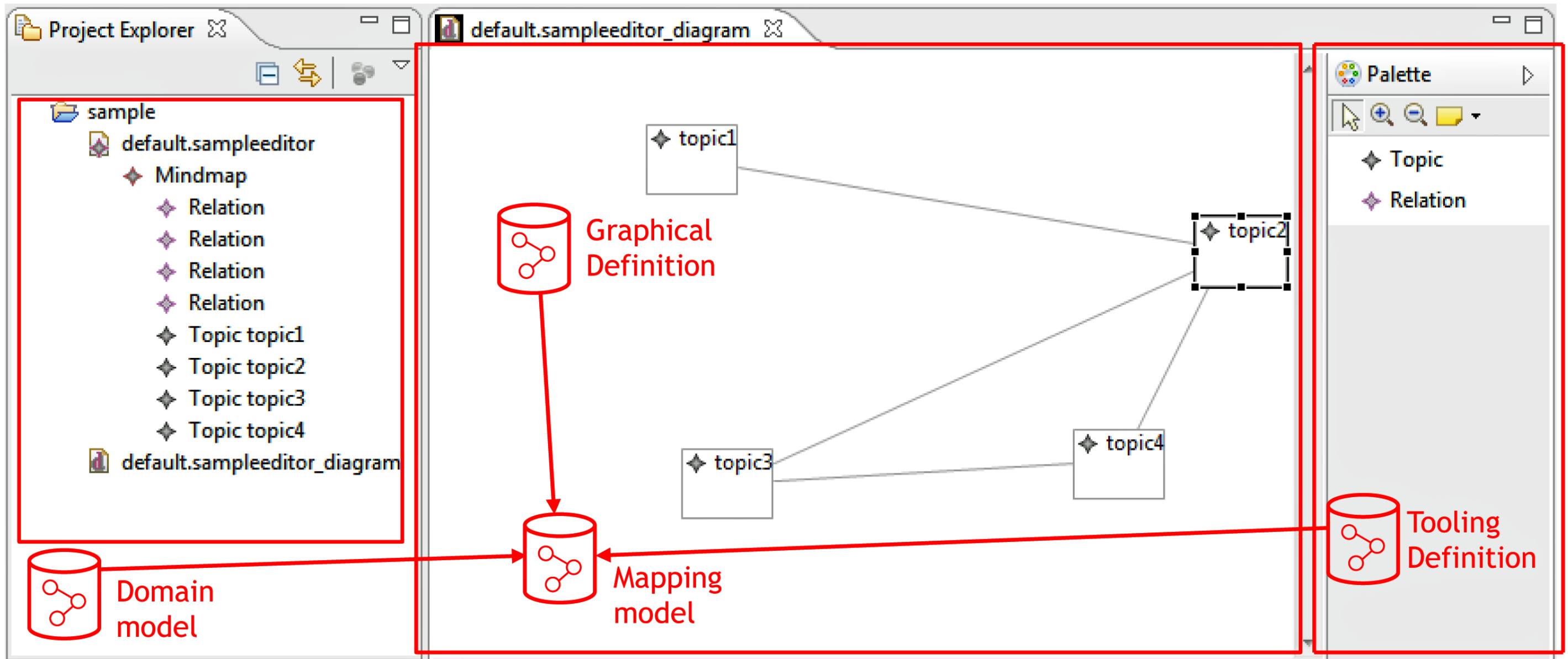}
\vspace{-55\in}
	\caption{Snapshot of a simple editor with indications of
          underlying models.}
	\label{F:editor}
\vspace{-42\in}
\end{figure}


The present paper describes research on the \emph{co-evolution} of GMF
editors. In particular, we are concerned with the question what and
how GMF models need to be co-changed in reply to changes of the domain
model (say, metamodel, or abstract syntax definition) of the
editor. Consider, for example, the simple mind-map editor in
Fig.~\ref{F:editor}.  We have annotated the different panes of the
editor with the associated GMF models underneath, and also added
mentioning of the mapping model which connects the various models. (We
will describe the architecture of the editor in more detail later.)
The co-evolution challenge at hand is to ``unbreak'' the editor upon
changes to the domain model. The GMF framework does not support such
co-evolution. In particular, there are no
semi-automatic means to unbreak the editor.

Such lack of co-evolution support somewhat diminishes the original
goal of GMF to aggressively simplify the development of graphical
editors. That is, while it is reasonable simple to draft and connect
all GMF models from scratch, it is notably difficult to evolve an
editor through changes of specific GMF models. A recurring focus for
evolution is the domain model of the editor. Upon domain model
changes, the user may notice that the editor is broken through
unsuccessful runs of some generator, the compiler, or the editor
itself, and in all cases, subject error messages at a low level of
abstraction. However, the editor's unsoundness may also go unnoticed
for some time. Alternatively, the user may attempt to regenerate some
models through the available wizards (model-to-model transformations
of GMF), which however means that the original, possibly customized
models are lost.


\vspace{-42\in}

\subsubsection*{Contributions}

\begin{itemize}

\item We analyze GMF's characteristics in terms of the co-evolution of
  the various models that contribute to a GMF editor. Starting from
  conceived domain-model changes, their implications for the editor
  itself and other GMF models are identified.

\medskip

\item We address the resulting co-evolution challenge by complementing
  GMF's wizard- and generator-biased architecture with model-to-model
  transformations that propagate changes of the domain model to
  other models.

\end{itemize}


\vspace{-42\in}

\subsubsection*{Limitations} The existing GMF infrastructure is
obviously rather complicated: it is a conglomeration of metamodels,
libraries, generators, model transformations of industrial scale. We
cannot claim to provide a full-fledged solution to the co-evolution
challenge of GMF---this would require full coverage of Ecore, and full
understanding of the implicit semantics of GMF model dependencies and
tools. Nevertheless, we are confident that our transformational
approach can be scaled incrementally over time to cover an increasing
number of concrete evolution scenarios. We make available some
reusable elements of our development publicly (scenarios,
transformations, difference models,
etc.).\footnote{\url{http://www.emfmigrate.org/gmfevolution}} The most
critical omission in our methodology is that we do not cover currently
co-evolution of customization code. This is a very intricate problem
by itself, to which we hope to contribute through future work.


\vspace{-42\in}

\subsubsection*{Road-map} In \S\ref{sec:background}, we briefly recall
the basics of the GMF approach to graphical editor development. In
\S\ref{sec:challenge}, we study a detailed evolution scenario to
systematically reveal the co-evolution challenge of GMF. In
\S\ref{sec:changesandco}, we develop an initial list of domain model
changes and derive a methodology of co-evolution based on propagating
changes to all relevant GMF models. In \S\ref{sec:transformations}, we
describe the realization of model-to-model transformations that are
driven by difference models for the domain-model changes. Related work
is described in \S\ref{sec:related}, and the paper is concluded in
\S\ref{sec:conclusions}.



\section{GMF in a nutshell}
\label{sec:background}

\vspace{-33\in}

\begin{figure}[t]
\vspace{-22\in}
	\centering
	\includegraphics[width=.97\linewidth]{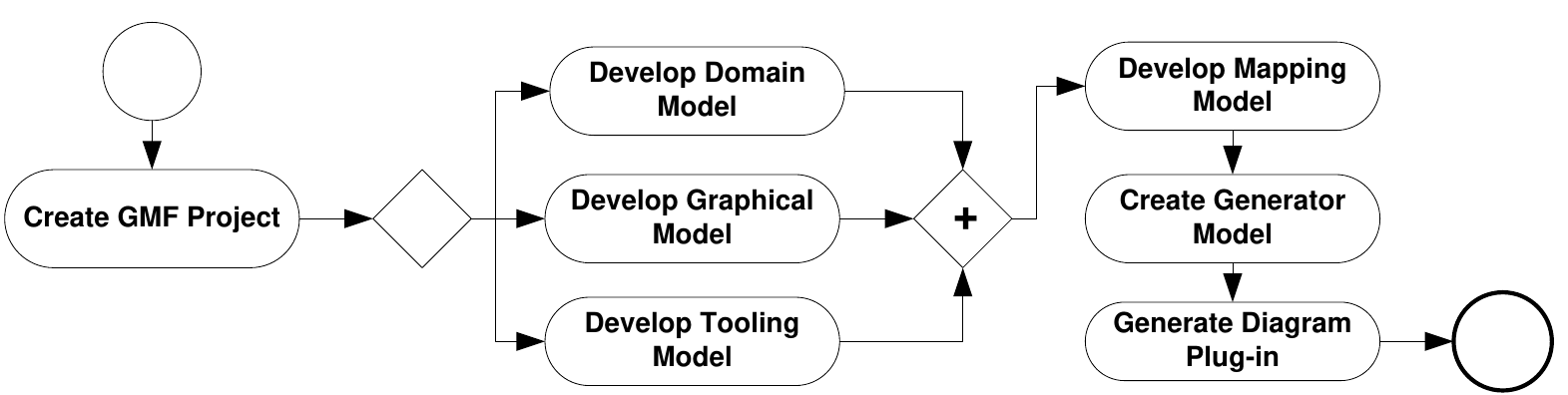}	
\vspace{-55\in}
	\caption{The model-driven approach to GMF-based editor development.}
	\label{F:process}
\vspace{-42\in}
\end{figure}

GMF consists of a generative component (GMF Tooling) and runtime
infrastructure (GMF Runtime) for developing graphical editors based on
the Eclipse Modeling Framework (EMF)~\cite{EMF1,EMF2} and the
Graphical Editing Framework (GEF)~\cite{GEF}. The GMF Tooling supports
a model-driven process (see Fig.~\ref{F:process}) for generating a
fully functional graphical editor based on the GMF Runtime starting
from the following models:
\begin{itemize}

\item The \emph{domain model} is the Ecore-based metamodel (say,
  abstract syntax) of the language for which representation and
  editing have to be provided.

\smallskip

\item The \emph{graphical definition model} contains part of the
  concrete syntax; it identifies graphical elements that may, in fact,
  be reused for different editors.

\smallskip

\item The \emph{tooling definition model} contains another part of the
  concrete syntax; it contributes to palettes, menus, toolbars, and
  other periphery of the editor.

\smallskip

\item Conceptually, the aforementioned models are reusable; they do
  not contain references to each other. It is the \emph{mapping model}
  that establishes all links.

\end{itemize}


\begin{figure}[th!]
\vspace{-42\in}
	\centering
	\includegraphics[width=.97\textwidth]{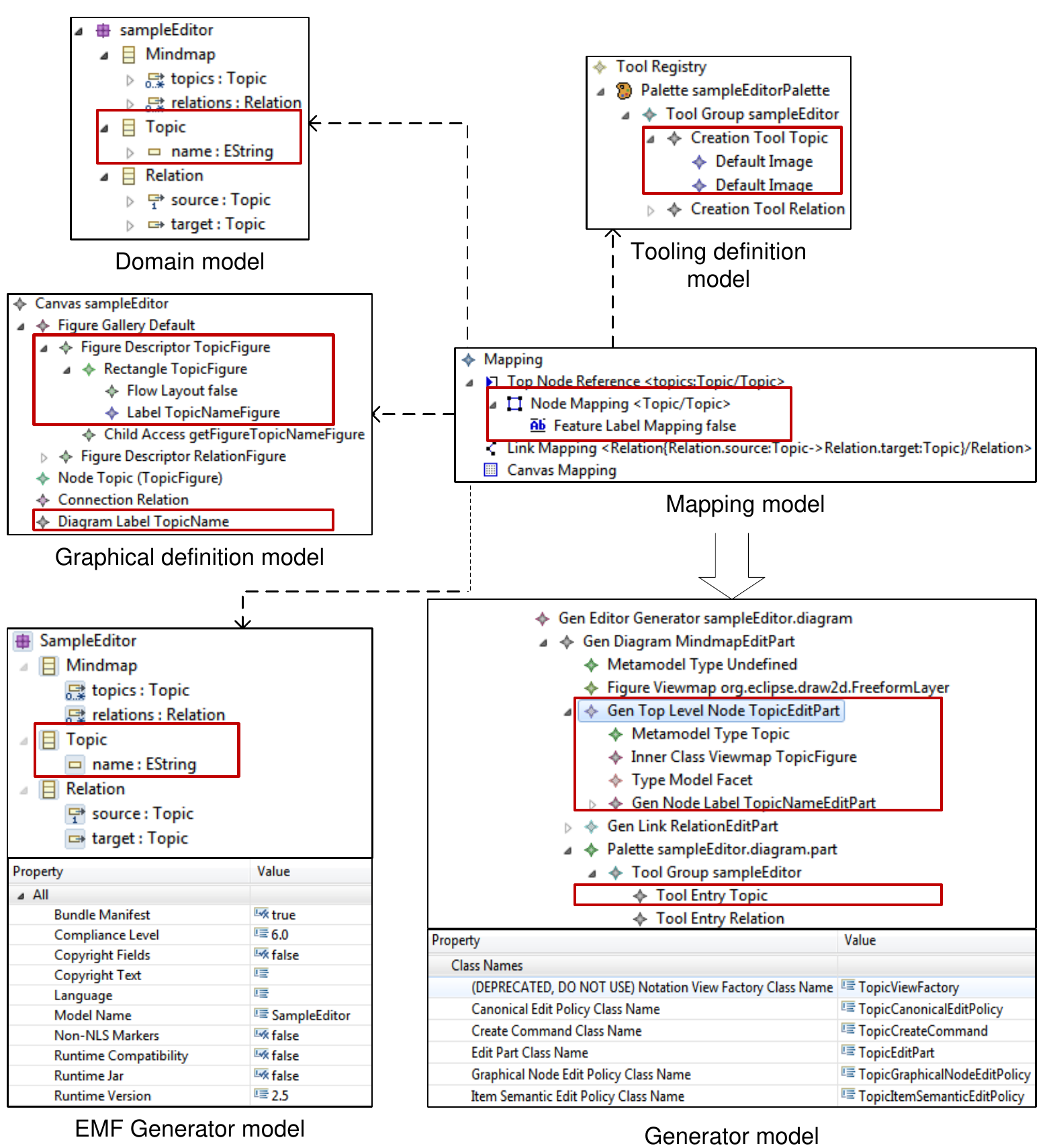}
\vspace{-77\in}
\caption{\textbf{The GMF models and model dependencies for the editor
    of Fig.~\ref{F:editor}.} We highlight contributions that are
  related to a selected domain concept: \emph{topics}. In this manner, we
  show how information about domain concepts is scattered over the
  various models. It is important to notice that most of these
  recurrences are not actual references; the correspondences are
  name-based. Strong links are only to be found in the mapping and the
  generator models.}
	\label{F:models}
\vspace{-66\in}
\end{figure}


Consider again Fig.~\ref{F:editor} which illustrates the role of these
models for a simple mind-map editor.\footnote{A mind map is a diagram
  used to represent words, ideas, tasks, or other items linked to and
  arranged around a central keyword or idea. The initial mind-map editor
  suffices with ``topics'' and ``relations'', but some
  extensions will be applied eventually.} Fig.~\ref{F:models} shows
all the models involved in the definition and implementation of the
mind-map editor. In addition to the aforementioned models, two
generator models are mentioned; they will be explained in a second.

The domain model of the mind-map editor contains all the concepts and
relationships which have to be implemented in the editor. In the
example, the class \emph{Mindmap} is introduced as a container of
instances of the classes \emph{Topic} and \emph{Relation}.

Once a domain model is defined, it is possible to automatically
produce Java code to manage models (instances), say mind maps in our
example. To this end an additional model, the \emph{EMF generator
  model}, is required to control the execution of the EMF generator. A
uniform version of the extra model can be generated by EMF
tooling. The model contains the mere list of classes and properties to
be considered as well as low-level details, e.g., the package prefix
for the generated code.

The graphical definition model consists of a figure gallery including
shapes, labels, lines, etc., and canvas elements to define nodes,
connections, compartments, and diagram labels. For instance, in the
graphical model in Fig.~\ref{F:models}, a rectangle named
\emph{TopicFigure} is defined, and it is referred to by the node
\emph{Topic}. A diagram label named \emph{TopicName} is also
defined. Such graphical elements will be used to specify the graphical
representations for \emph{Topic} instances and their \emph{name}s.

The tooling definition model defines toolbars and other periphery to
facilitate the management of diagram content. In Fig.~\ref{F:models},
the sample model consists of the \emph{Topic} and \emph{Relation}
tools for creating the \emph{Topic} and \emph{Relation} elements.

The mapping model links together the various models. For instance,
according to the mapping model in Figure~\ref{F:models}, \emph{Topic}
elements are created by means of the \emph{Creation Tool Topic} and
the graphical representation for them is \emph{Node Topic}. For each
topic the corresponding \emph{name} is also visualized because of the
specified \emph{Feature Label Mapping} which relates the attribute
\emph{name} of the class \emph{Topic} with the diagram label
\emph{TopicName} defined in the graphical definition model.

Once the mapping model is obtained, the GMF Tooling generates (by
means of a model-to-model transformation) the \emph{GMF generator
  model} that is used by a code generator to produce the real code of
the modeled graphical editor.

\vspace{-42\in}

\section{GMF's co-evolution challenge}
\label{sec:challenge}

\vspace{-33\in}

Using a compound change scenario, we will now demonstrate GMF's
co-evolution challenge. Overall, the objective is to clarify that
domain model changes imply changes of other GMF models. Such change
propagation is not supported currently by GMF, and it is
labor-intensive and error-prone, when carried out
manually. Conceptually, it turns out to be difficult to precisely
predict when and how co-changes must be performed.


\begin{figure}[h!]
\vspace{-55\in}
	\centering
		\includegraphics[width=.82\linewidth]{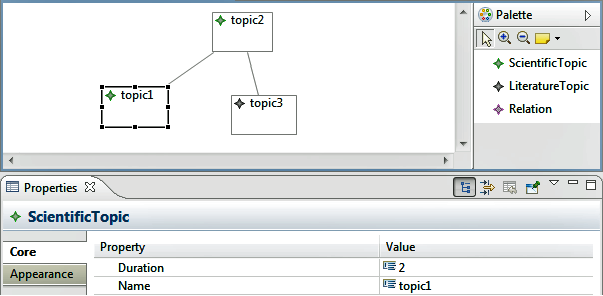}
\vspace{-42\in}
	\caption{An evolved mind-map editor with different kinds of topics.}
	\label{F:evolvedEditor}
\vspace{-42\in}
\end{figure}


Consider the enhanced mind-map editor of
Fig.~\ref{F:evolvedEditor}. Compared to the initial version of
Fig.~\ref{F:editor}, \emph{scientific} vs.\ \emph{literature topics} are
distinguished, and topics have a \emph{duration} property in addition to the
\emph{name} property.


\begin{figure}[t!]
\vspace{-42\in}
\centering
\fbox{\includegraphics[width=10cm]{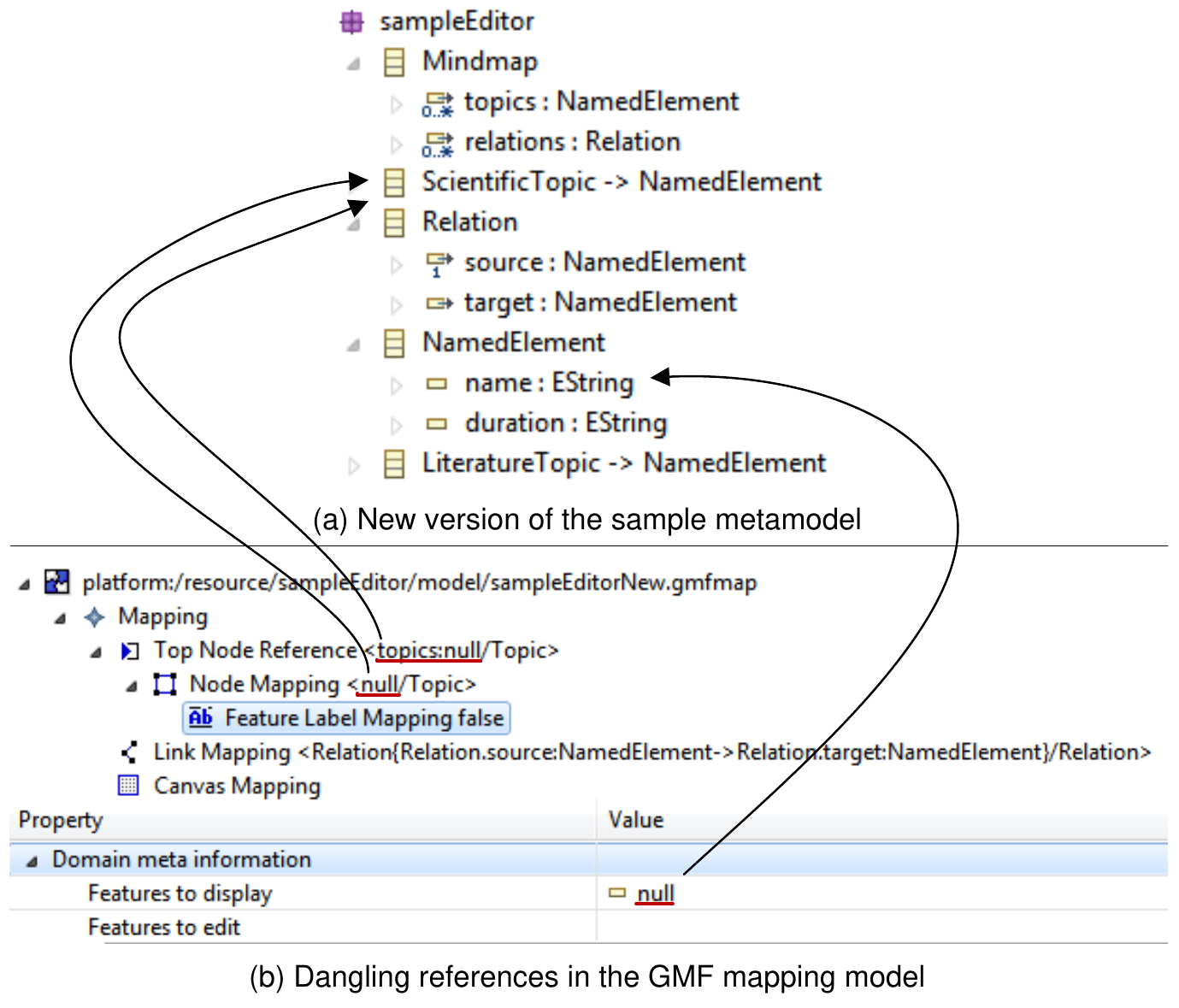}}
\caption{The domain model for the evolved mind-map editor with the ``broken'' mapping model.}
\label{F:modelMismatch}
\vspace{-42\in}
\end{figure}


Now consider Fig.~\ref{F:modelMismatch}; it shows the evolved
metamodel at the top, and the status of the, as yet, unamended mapping
model at the bottom. We actually show the mapping model as it would
appear to the user if it was inspected in Eclipse. Some of the links
in the mapping model are no longer valid; in fact, they are dangling
(c.f., ``null''). Through extra edges, we show what the links are
supposed to be like.

We get deeper insight into the situation if we comprehend the evolved
domain model through a series of simple, potentially atomic
changes. i) The \emph{Topic} class was renamed to
\emph{ScientificTopic}. ii) The new abstract class \emph{NamedElement}
was added. iii) The attribute \emph{name} of the old \emph{Topic}
class was pulled up to the new \emph{NamedElement} class. iv) The
attribute \emph{duration} was added to the \emph{NamedElement}
class. v) The new class \emph{LiteratureTopic} was added as a further
specialization of \emph{NamedElement}.

In practice, these changes would have been carried out in an ad-hoc
manner through free-wheeling editing capabilities of the Ecore editor
of Eclipse. Because of these changes, the existing mapping model is no
longer valid---as shown in Fig.~\ref{F:modelMismatch}. In particular,
references to \emph{Topic} or the attribute \emph{name} thereof are
dangling. The other GMF models are equally out-of-sync after these
domain model changes.

Incomplete or unsuccessful editor evolution may be signaled in
different ways, or may, in fact, remain hidden for some
time---depending on the specific domain model change as well as the
usage of GMF tooling. We list obvious and hidden symptoms of broken or
unsound editors:

{\small

\begin{enumerate}

\item The EMF generator or the GMF generator fails (with an error).
\item The EMF generator or the GMF generator completes with a warning.
\item The generator for the GMF generator model fails.
\item The compiler fails on the generated EMF or GMF code.
\item The editor plugin fails at runtime, e.g., at launch-time.
\item A GMF model editor reports an error upon opening a GMF model.
\item The editor plugin apparently executes but misses concepts of the domain.
\item The editor plugin apparently executes but there are GUI events without handlers.

\end{enumerate}

}

Let us consider two specific examples. First, the addition of a new
class to the domain model, e.g., \emph{LiteratureTopic}, should
probably imply a capability of the editor to create instances of the
new class. However, such a creation tool would need to be added in the
mapping and tooling models. Second, the renaming of a class, e.g., the
renaming of \emph{Topic} to \emph{ScientificTopic}, may lead to an
editor with certain functionality not having any effect because
elements are referenced that changed or do not exist anymore in the
domain model. Both examples are particularly interesting in so far
that the editor apparently still works. i.e., it is not \emph{broken}
in a straightforward sense. However, we say that the editor is
\emph{unsound}; the editor does not meet some obvious expectations.

Such a distinction of \emph{broken} vs.\ not broken but nevertheless
\emph{unsound} also naturally relates to a spectrum of
\emph{strategies} for co-changes. A \emph{minimalistic strategy} would
focus on unbreaking the editor. That is, co-changes are supposed to
bring the editor models to a state where no issues are reported at
generation, compile or runtime. A \emph{best-effort strategy} would
try to bring the editor to a sound state, or as close to it as
possible with a general (perhaps automated) strategy.

Consider again the example of adding a new class $C$:

\vspace{-22\in}

{\small 

\begin{description}

\item[Minimalistic strategy] The execution of the EMF generator emits
  a warning, which we take to mean that the editor is broken. Hence,
  we would add the new class $C$ to the EMF generator model. This
  small co-change would be sufficient to re-execute all generators
  without further errors or warnings, and to build and run the editor
  successfully. The editor would be agnostic to the new class though
  because the mapping and tooling models were not co-changed.

\item[Best-effort strategy] Let us make further assumptions about the
  added class $C$. In fact, let us assume that $C$ is a concrete
  class, and it has a superclass $S$ with at least one existing
  concrete subclass $D$. In such a case, we may co-change the other
  GMF models by providing management for $C$ based on the replication 
  of the management for $D$.

\end{description}

}

\vspace{-22\in}

Here we assume that a best-effort strategy may be amenable to an
automated transformation approach in that it does not require any
domain-specific insight. The modeler will still need to perform
additional changes to complete the evolution, i.e., to obtain a sound
editor.


\vspace{-42\in}

\section{Changes and co-changes}
\label{sec:changesandco}

\vspace{-33\in}

Let us discuss and assess a catalog of domain-model changes and
associated co-changes of other editor models. Obviously, we can depart
from catalogs of metamodel changes as they are available in the
literature, e.g., \cite{Wachsmuth07,HerrmannsdoerferBJ09}, and
previous work by the authors~\cite{CicchettiDREP08}. Due to space
constraints, a selection must be made: it covers atomic changes that
are needed for the compound scenario of the previous section,
completed by
a few additional ones. Many of the missing changes would refer
to technical aspects of the EMF implementation, and as such, they do
not contribute to the discussion.


\begin{table}[h!]
\vspace{-22\in}
	\centering
		\begin{tabular}{|c||l|}\hline
                  \ Level \ & Description \\\hline\hline
                  1 & Unsound in the sense of being broken; there are reported issues (errors, warnings). \\\hline
                  2 & Unsound in the sense that the editor ``obviously'' lacks
                  capabilities. \\\hline
                  3 & Sound as far as it can be achieved through automated transformations.\\\hline
                  4 & Sound; established by human evaluation. \\\hline
						
		\end{tabular}
\smallskip
	\caption{Levels of editor soundness along evolution.}
	\label{T:levels}
\vspace{-99\in}
\end{table}

In continuation of the soundness discussion from the previous section,
Table~\ref{T:levels} identifies different levels of soundness for an
evolving editor. The idea is here that we assess the level of the
editor \emph{before} and \emph{after} all (automated) co-changes were
applied. The proposed transformations can never reach Level 4 because
it requires genuine evaluation by the modeler.
 

\begin{table}[t]
\vspace{-52\in}
\centering

{\footnotesize

\begin{tabular}{|l||c|c|c|c|c||c|c|c|c|c|}
\hline
& \multicolumn{5}{|c||}{\begin{tabular}{c}before\\co-change\end{tabular}} 
& \multicolumn{5}{|c|}{\begin{tabular}{c}after\\co-change\end{tabular}} 
\\ \hline
 & \sw{EMFGen\ \ } & \sw{Graph} & \sw{Tooling} & \sw{Mapping} & \sw{Level}
 & \sw{EMFGen} & \sw{Graph} & \sw{Tooling} & \sw{Mapping} & \sw{Level}\\ \hline\hline
Add empty, concrete class
& \broken & \unsound  & \unsound  & \unsound & 1
& \sound & \unsound  & \unsound  & \unsound & 2
\\ \hline 
Add empty, abstract class
& \broken & \sound  & \sound  & \sound & 1
& \sound & \sound  & \sound  & \sound & 3
\\ \hline 
Add specialization
& \sound & \sound  & \sound  & \sound  & 3
& \sound & \sound  & \sound & \sound & 3
\\ \hline 
Delete concrete class
& \broken & \unsound  & \broken  & \broken  & 1
& \sound & \unsound  & \sound & \sound & 2
\\ \hline 
Rename class 
& \broken & \unsound  & \unsound  & \broken  & 1
& \sound & \sound  & \sound & \sound & 3
\\ \hline 
Add property
& \broken & \unsound  & \unsound  & \unsound & 1
& \sound & \unsound  & \unsound  & \sound & 2
\\ \hline
Delete property
& \broken & \unsound  & \broken  & \broken  & 1
& \sound & \unsound  & \sound & \sound & 2
\\ \hline
Rename property
& \broken & \unsound  & \unsound  & \broken  & 1
& \sound & \sound  & \sound & \sound & 3
\\ \hline
Move property
& \broken & \unsound  & \broken  & \broken  & 1
& \sound & \unsound  & \unsound & \unsound & 2
\\ \hline 
Pull up property
& \broken & \unsound  & \broken  & \broken  & 1
& \sound & \unsound  & \sound & \sound & 2
\\ \hline
Change property type
& \sound & \unsound  & \broken  & \broken  & 1
& \sound & \unsound  & \unsound & \unsound & 2
\\ \hline  
\end{tabular}
}

\smallskip

\caption{Considered Ecore metamodel changes}
\label{F:blame}
\vspace{-99\in}
\end{table}


However, we are not just interested in the overall level of the
editor, but we also want to \emph{blame} one or more editor models for
the editor's unsoundness. In Table~\ref{F:blame}, we list atomic
changes with the soundness levels for the editor before and after
co-changes, and all the indications as to what models are to blame. We
use ``\broken'' to blame a model for causing the editor to be broken,
i.e., to be at Level 1. We use ``\unsound'' and ``\sound'' likewise
for Level 2 and Level 3.

The EMFGen model is frequently to blame for a broken editor \emph{before}
the co-changes; the Graph model is never to blame for a broken editor;
the remaining models are to blame occasionally for a broken
editor. Obviously, there is trend towards less blame after
the co-changes: no occurrences of ``\broken'', more occurrences of
``\sound''.  In different terms, for all domain-model changes, all
other models can be co-changed so that the editor is no longer
broken. In several cases, we reach Level 3 for the editor.

There are clearly constellations for which changes cannot be
propagated in an automated manner that resolves all Level 2 blame. For
instance, the metamodel change \emph{add concrete class} does not
require a co-changed Graph model as long as some existing graphical
element can be reused. However, avoidance of Level 2 blame would
require a manual designation of a new element or genuine selection of
a specific element as opposed to an automatically chosen element.

Fig.~\ref{F:details} describes some changes and co-changes in detail.

\begin{figure}[t!]
\vspace{-33\in}
\begin{boxedminipage}{\hsize}

\noindent\emph{Add empty, concrete class} Apart from the EMFGen model,
the other ones are not affected; the editor simply does not take into
account the added class. Thus, modelers cannot create or edit
instances of the new class. The co-change may replicate the model from
existing classes as discussed in \S\ref{sec:challenge}. Ultimately, the
modeler may need to manually complete the management of the new class.

\smallskip

\noindent\emph{Add empty, abstract class} In comparison to the previous case,
the co-change of the EMFGen model is fully sufficient since
abstract classes cannot be instantiated, and hence, no additional
functionality is needed in the editor.

\smallskip

\noindent\emph{Add specialization} The change consists of modifying an
existing class by specifying it as specialization of another one. In
particular, in the simple case of the superclass being empty, this
modification does not affect any model; thus, no co-changes are
required.

%

\smallskip

\noindent\emph{Delete concrete class} Deleting an existing class
is more problematic since all the GMF models except the Graph model
are affected. Especially the Mapping model has to be fixed to solve
possible dangling references to the deleted class. The Tooling model
is also co-changed by removing the creation tool used to create
instances of the deleted class. Even if the model is not adapted, the
generator model and thus the editor can be generated---even though the
palette will contain a useless tool without associated
functionality. The Graph model can be left unchanged. The graphical
elements which were used for representing instances of the deleted
class, may be re-used in the future.

\smallskip

\noindent\emph{Rename class} Renaming a class requires co-change of
the Mapping model which can have, as in the case of class deletion,
invalid references which have to be fixed by considering the new class
name. The Graph model does not require any co-change since the
graphical elements used for the old class can be used even after the
rename operation. The Tooling model can be left untouched, or
alternatively the label and the description of the tool related to the
renamed class can be modified to reflect the same name. However, even
with the same Tooling model, a working editor will be generated.

\smallskip

\noindent\emph{Add property} The strategy for co-change is similar to the
addition of new classes.


\smallskip

\noindent\emph{Delete property} Deleting a property which
has a diagrammatic representation requires a co-change of the Mapping
model in order to fix occurred dangling references. Moreover if some
tools were available to manage the deleted property, also the Tooling
model has to be co-changed. As in the case of class removals, the
graphical model can be left unchanged.

\smallskip

\noindent\emph{Rename property} The strategy for co-change is similar to the
renaming of classes.


\smallskip

\noindent\emph{Move property} When a property is moved from one
class to another, then dangling references may need to be resolved in
the Mapping model. If the moved property is managed by means of
some tools, the Tooling model require co-changes, too. We only offer a
simple, generic strategy for co-changes: the repaired editor does not
consider the moved property.

\smallskip

\noindent\emph{Pull up property} Given a class hierarchy, a given
property is moved from an extended to a base class. This modification
is similar to the previous one---even though an automatic resolution can
be provided to co-change Tooling and Mapping models in a satisfactory
manner.

\smallskip

\noindent\emph{Change property type} The EMFGen model is not affected.
However, by changing the type of a property some dangling references
can occur in the Mapping model; their resolution cannot be fully
automated. Also, if the affected property is managed by some tool,
then the Tooling model must be co-changed as well.

\end{boxedminipage}
\caption{Detailed descriptions of selected changes and co-changes.}
\label{F:details}
\vspace{-42\in}
\end{figure}


\begin{figure}[t]
	\centering
		\includegraphics[width=11cm]{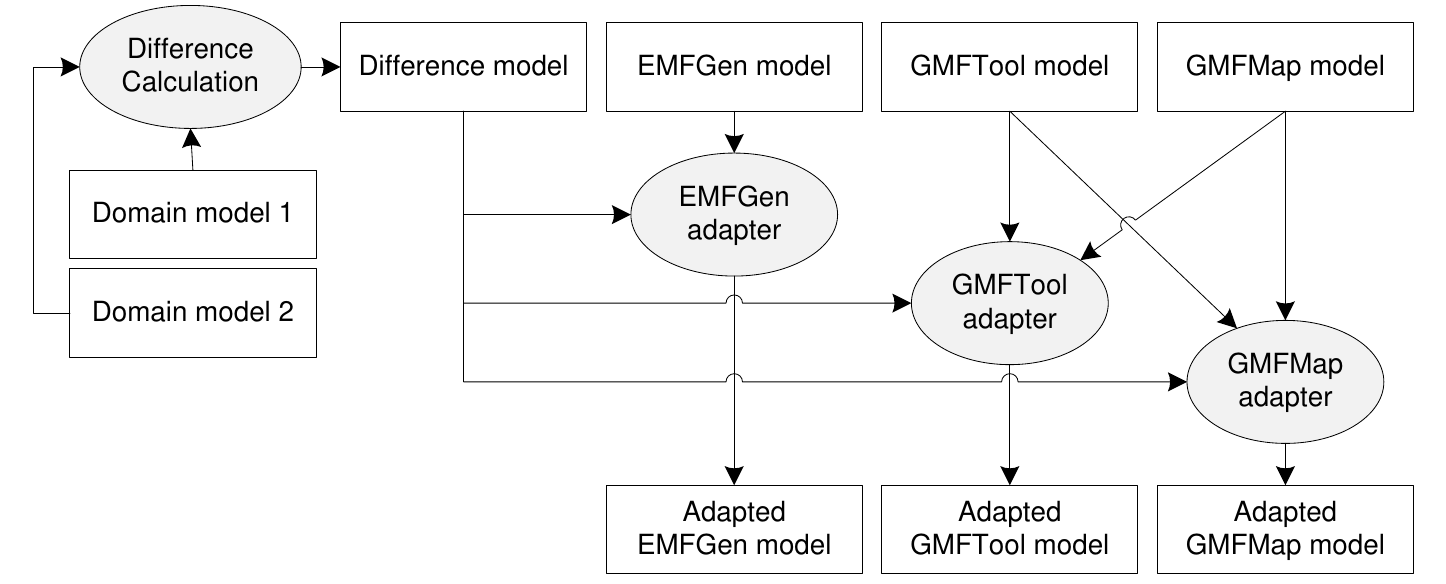}
\vspace{-33\in}
	\caption{Overview of the process of co-evolution with automated transformations.}
	\label{F:adapters}
\end{figure}


\section{Automated transformation of GMF models}
\label{sec:transformations}

We have developed a process for GMF co-evolution which involves
automated transformations in an essential manner. We describe this
process here. We also provide some insight into the implementation of
the involved transformations, which is based on model transformations
specified in ATL~\cite{JouaultK05}. (The implementation is available
publicly---as described in the introduction of the paper.)

Fig.~\ref{F:adapters} summarizes the approach: given two subsequent
versions of the same \emph{domain model} their differences are
calculated and then represented in a \emph{difference model}. Such
differences are then taken as input by different adapters each devoted
to the co-change of a specific GMF model. Interestingly, the GMFMap
and the GMFTool adapters take as input both the \emph{GMFMap model}
and the \emph{GMFTool model} since the dependencies between these two
models have to be updated simultaneously. No adapter is
provided for the Graph model; the discussion of the previous sections
suggested that we can always reasonable continue with the old model.


\begin{figure}[b]
  \centering
  \includegraphics[width=1\textwidth]{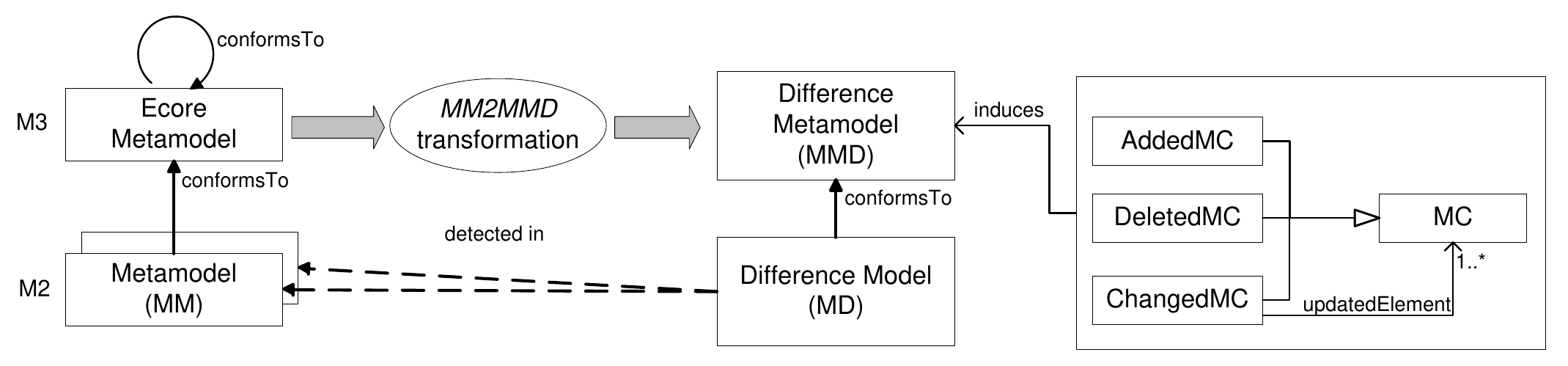}
\vspace{-52\in}
  \caption{Difference metamodel generation}
  \label{F:diff-generation}
\vspace{-33\in}
\end{figure}


\vspace{-22\in}

\subsection*{Model-based representation of domain model differences}

We seek the adoption of standard model-driven techniques and tools for
their management and to derive in an automatic way the co-changes of
the GMF models. To this end, the \emph{difference metamodel} concept,
presented by the authors in~\cite{CicchettiDP07}, and already adopted
in different scenarios including the management of the coupled
evolution of metamodels and models~\cite{CicchettiDREP08}, is
leveraged. 

The approach is summarized in Fig.~\ref{F:diff-generation}:
given two Ecore metamodels, their difference conforms to a difference
metamodel \emph{MMD} derived from Ecore by means of the \emph{MM2MMD}
transformation. For each class \emph{MC} of the Ecore metamodel,
the additional classes \emph{AddedMC}, \emph{DeletedMC}, and
\emph{ChangedMC} are generated in the extended Ecore metamodel by
enabling the representation of the possible modifications that can
occur on domain models and that can be grouped as follows:

\begin{itemize}
	\item[--] \emph{additions}, new elements are added in the initial metamodel; 
	\item[--] \emph{deletions}, some of the existing elements are deleted;
	\item[--] \emph{changes}, some of the existing elements are updated. 
\end{itemize}

\begin{figure}[t]
\vspace{-33\in}
  \centering
  \includegraphics[width=1\textwidth]{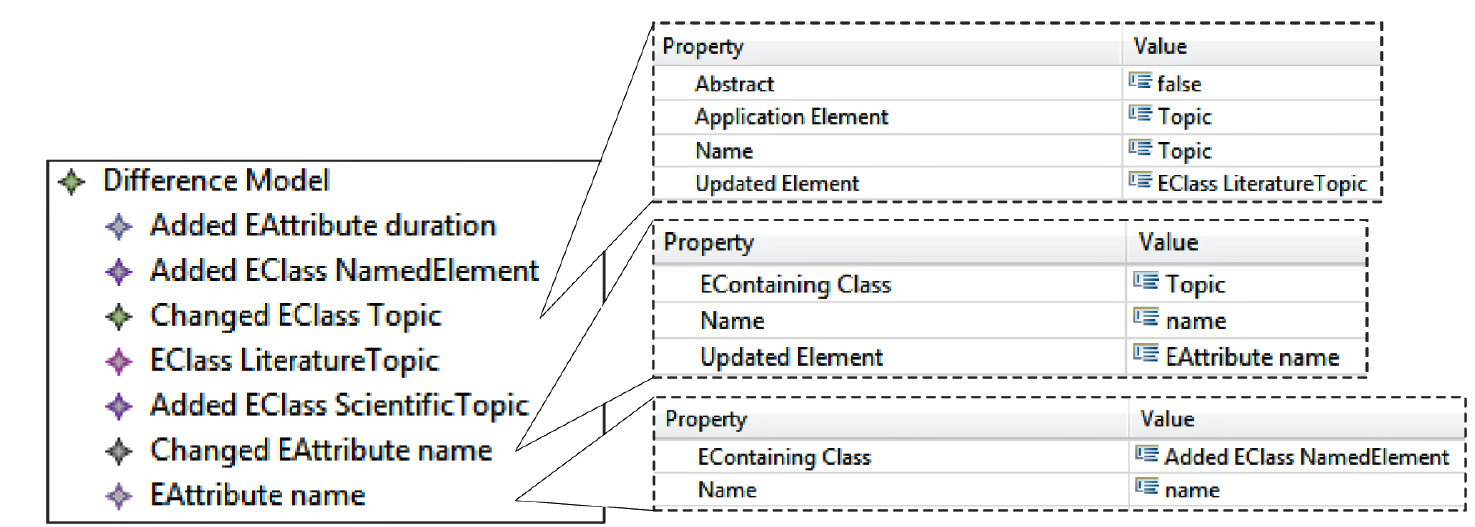}
\vspace{-66\in}
  \caption{Fragment of the difference model for the evolution scenario of \S\ref{sec:challenge}.}
  \label{fig:sampleDifferenceModel}
\vspace{-42\in}
\end{figure}

In Fig.~\ref{fig:sampleDifferenceModel}, a fragment of the difference
model representing the changes between the domain models in
Fig.~\ref{F:models} and Fig.~\ref{F:modelMismatch} is
shown. For some of the reported differences, the corresponding
properties are shown. For instance, the renaming of the \emph{Topic}
class is represented by means of a \emph{ChangedEClass} instance which
has as updated element an instance of \emph{EClass} named
\emph{LiteratureTopic} (see the \emph{updatedElement} property of the
changed class \emph{Topic} shown on the right-hand side of
Fig.~\ref{fig:sampleDifferenceModel}). The addition of the class
\emph{NamedElement} is represented by means of an \emph{AddedEClass}
instance. The move operation of the attribute \emph{name} from the
class \emph{Topic} to the added class \emph{NamedElement} is
represented by means of a \emph{ChangedEAttribute} instance which has
one \emph{EAttribute} instance as updated element with a different
value for the \emph{eContainingClass} property. In fact, in the
initial version it was \emph{Topic} (see the second property window)
whereas in the last one, it is \emph{NamedElement} (as specified in
the third property window).


\vspace{-22\in}

\subsection*{ATL-based implementation of GMF model adapters}

Our prototypical implementation of the GMF model adapters leverages
ATL, a QVT compliant language which contains a mixture of declarative
and imperative constructs. In particular, the implementation of the
GMF adapters consists of transformation rules which copy the given
source model to a target one; during this operation they evaluate if
changes are needed. A number of helpers have been defined; they
navigate models and perform complex queries on them. Many of the
helpers are common to all the adapters, and hence, they are available
through a library \emph{gmfAdaptationLib}. Table~\ref{tab:helpers}
describes some of these helpers.


\begin{table}\footnotesize
\vspace{-42\in}
	\centering
		\begin{tabular}{|c|c|c|p{5.3cm}|}
    \hline
    {\bf Helper name} & {\bf Context} & {\bf Return type} & {\bf Description}\\
    \hline
    getEClassInNewMetamodel & EClass & EClass &  Given a class of the old metamodel, it returns the corresponding one in the new
metamodel. \\
    \hline
getNewContainer & EAttribute & EClass & Given an EAttribute in the old metamodel, the corresponding container in the new one is retrieved. To this end, the helper checks if the EAttribute has been moved to a new added class, if not an existing class is returned.\\
    \hline
    isMoved & EAttribute & Boolean & It checks if the considered EAttribute has been moved to another container \\
    \hline
    isMovedToAddedEClass & EAttribute & Boolean & It checks if the considered EAttribute has been moved to a new added EClass.\\
    \hline
    isRenamed & EAttribute & Boolean & It checks if the given EAttribute has been renamed.\\
    \hline
		\end{tabular}
\caption{Some helpers of the \emph{gmfAdaptationLib}}
\label{tab:helpers}
\vspace{-66\in}
\end{table}


Small excerpts of the GMFMap and GMFTool adapters are shown in
Listing~\ref{lst:GMFMapAdapter} and Listing~\ref{lst:GMFToolAdapter},
respectively.  For instance, the \emph{AddedSpecializationClassTo...}
transformation rules manage new classes which have been added in the
domain model as specializations of an existing one. The code excerpts
involve the replication strategy that we have described in previous
sections. More specifically, the
\emph{AddedSpecializationClassToNodeMapping} rule in
Listing~\ref{lst:GMFMapAdapter} is executed for each match of the
source pattern in lines 4-14 which describes situations like the one
we had in the sample scenario where the \emph{LiteratureTopic} class
(see \emph{s1}) is added as specialization of an abstract class (see
\emph{s2}) which is specialized by another class (see \emph{s3}). In
this case, the Mapping model is updated by adding a new
\emph{TopNodeReference} and its contained elements (see lines 20-34)
which are copies of those already existing for \emph{s3}.


\medskip

\begin{lstlisting}[breaklines,style=AMMA,language=ATL,mathescape,rulesepcolor=\color{black},caption=Fragment
of the GMFMap Adapter,captionpos=b,label={lst:GMFMapAdapter}]
...
rule AddedSpecializationClassToNodeMapping {
	
  from 
   s1: DELTAMM!AddedEClass, s2: DELTAMM!AddedEClass,	
		 s3: DELTAMM!ChangedEClass, s4: DELTAMM!ChangedEAttribute,
		 s5: DELTAMM!EAttribute
	  	 ((not s1.abstract) 
      and s1.eSuperTypes->first() = s2 
      and s2.abstract 
   			and s3.updatedElement->first().eSuperTypes->first() = s2
   			and s4.updatedElement->first() = s5 
      and s4.eContainingClass = s3 
      and s5.eContainingClass = s2 ))
 	
  using {
	  	siblingFeatureLabelMapping : GMFMAPMM!FeatureLabelMapping = s3.getNodeMappingFromChangedClass().labelMappings
													->select(e | e.oclIsTypeOf(GMFMAPMM!FeatureLabelMapping))->first();	}
	
  to 
   t1 : GMFMAPMM!TopNodeReference (
 	  	 containmentFeature <- s3.getTopNodeReferenceFromChangedClass().containmentFeature.getFeatureInNewMetamodel(),
 	  	 ownedChild <- t2
    ),
   t2 : GMFMAPMM!NodeMapping (
 	  	 domainMetaElement <- s1.getAddedClassInNewMetamodel(),
    	 relatedDiagrams <- s3.getNodeMappingFromChangedClass().relatedDiagrams,
 	  	 tool <- s1.name.getNewToolFromTitle(),
 	  	 diagramNode <- s3.getNodeMappingFromChangedClass().diagramNode
    ),	 
   t3 : GMFMAPMM!FeatureLabelMapping (
 	  	 diagramLabel <- siblingFeatureLabelMapping.diagramLabel,
 	  	 features <- siblingFeatureLabelMapping.features->collect(e | e.getFeatureInNewMetamodel())
    ),
    ...
}
\end{lstlisting}


\medskip

A similar source pattern is used in the rule of
Listing~\ref{lst:GMFToolAdapter} (lines 4-9) in order to add a
creation tool for the new added class \emph{s1} to the Tooling model
(see lines 15-19).

\medskip


\begin{lstlisting}[breaklines,style=AMMA,language=ATL,mathescape,rulesepcolor=\color{black},caption=Sample transformation rule 
of the GMFTool Adapter,captionpos=b,label={lst:GMFToolAdapter}]
...
rule AddedSpecializationClassToCreationTool {
	
  from 
   s1: DELTAMM!AddedEClass, s2: DELTAMM!AddedEClass, s3: DELTAMM!ChangedEClass  
		  ( (not s1.abstract) 
      and s1.eSuperTypes->first() = s2
	   		and s2.abstract 
      and s3.updatedElement->first().eSuperTypes->first() = s2 )
	
  using {
	 	toolGroup : GMFTOOLMM!ToolGroup = OclUndefined;	
 	}
	
  to 
   t : GMFTOOLMM!CreationTool (
	    title <- s3.getToolFromChangedClass().title.regexReplaceAll(s3.getToolFromChangedClass().title, s1.name),
		   description <- 'Create new ' + s1.name
     ),
	    ...
}
\end{lstlisting}



\vspace{-42\in}

\section{Related work}
\label{sec:related}


\subsubsection*{Graphical model editors}

\vspace{-22\in}

In \cite{AmyotFR06}, a number of technologies for the development of
domain-specific modeling languages (DSMLs) are evaluated; Eclipse (EMF
with GEF) is covered, but not GMF. The evaluation criteria include
language evolution to mean the ability to co-evolve models when the
domain model changes, and Eclipse receives a medium grade here. There
is no criterion though that relates to GMF's particular
characteristics of using multiple editor models.

Other GMF- or GEF-based frameworks have been proposed. For instance,
the MuvitorKit framework~\cite{ModicaBE09} is based on EMF and GEF and
specifically meant as an alternative to GMF for the benefit of
additional editor capabilities (e.g., multiple panes) as well as
additional modeling capabilities, thereby requiring less customization
of generated code. There is also the EuGENia
framework~\cite{KolovosRPP09} which raises the level of abstraction in
GMF-based development by using annotations on the domain model,
thereby feeding into code generation. We are not aware of any prior
effort to propagate changes across GMF models.

The ViatraDSM framework~\cite{RathOV09} replaces GMF in that it allows
for versatile mappings between abstract and concrete syntax.  Live
transformations are leveraged to maintain the coherence of the two
models. Our uni-directional, difference-driven transformations
propagate domain-model changes elsewhere. Our work is specifically
targeted at the mainstream GMF-based approach with its various models.


\vspace{-33\in}

\subsubsection*{Model consistency}

\vspace{-22\in}

The status of GMF models being out-of-sync can be compared to the
notion of model inconsistency in (UML-based) modeling where different
models providing different views may require synchronization. For
instance, in \cite{EgyedLF08}, inconsistencies between the different
diagrammatic forms in UML models are considered, and possible fixes
are proposed in the form of value changes. In \cite{GieseW06}, the
dependencies between models are modeled through triple graph
grammars in a manner that enables incremental model
synchronization. Our specific contribution is one of reverse
engineering: discovering the GMF model dependencies, and making them
operational through automated transformations.


\vspace{-33\in}

\subsubsection*{Co-evolution of metamodels and models}

\vspace{-22\in}

{\sloppypar

  The techniques and the methodology of our work are inspired by
  research on co-evolution in model-driven
  engineering~\cite{Favre03,VermolenV08}. Much of this work is
  concerned with co-transforming models in reply to metamodel changes.
  In our work, the focus is on the domain models (metamodels), too,
  and changes are to be propagated to other editor models. Our
  soundness levels for evolved editors relate to transformation
  properties of~\cite{GruschkoKP07,Wachsmuth07}. Our change scenarios
  are inspired by considerations in
  \cite{Wachsmuth07,HerrmannsdoerferBJ09} for MOF and EMF. We leverage
  difference representations of our previous
  work~\cite{CicchettiDREP08,CicchettiRP10}.

}


\vspace{-22\in}

\section{Concluding remarks}
\label{sec:conclusions}

\vspace{-22\in}

We have described the challenge of sound evolution for graphical
editors based on model-driven development with GMF in particular, and
we have addressed this challenge by a system of co-transformations
that propagate changes from domain models to the other editor models.

We have identified a range of options for evolved editors to be
unsound, and we have described corresponding resolution
strategies. Our work is the first attempt to come to a similar level
of understanding as with the established problem of metamodel/model
co-evolution, in which case the situation is more clear-cut: models
either are not broken, or they are broken and can be reasonably
resolved in an automated manner, or a well-understood problem-specific
contribution to the resolution must be provided manually or through a
heuristic. In the case of co-evolution for editor models, each of the
various models calls for a designated analysis, and there are
intricate inter-model dependencies.

In our ongoing research, we try to better understand the co-evolution
issues and associated strategies for the code level of GMF where
generated code has been possibly customized. Based on preliminary
research, we can already report that customization is used by some GMF
projects extensively, and hence designated co-evolution support may
provide significant help with real-world editor development.


{\footnotesize

\bibliographystyle{abbrv}
\bibliography{paper}

}

\end{document}